\documentclass[11pt,a4paper]{article}
\usepackage{amsfonts,amssymb,amsmath,amsthm,cite}
\usepackage{graphicx}
\usepackage{slashed}
\usepackage[applemac]{inputenc}

\evensidemargin=\oddsidemargin
\DeclareMathOperator{\Dom}{Dom}      
\DeclareMathOperator{\End}{End}        
\DeclareMathOperator{\Id}{Id}                 
\DeclareMathOperator{\Ker}{Ker}           
\DeclareMathOperator{\Tr}{Tr}                 

\newtheorem{assumption}{Assumption}[section]
\newtheorem{theorem}[assumption]{Theorem}
\newtheorem{corollary}[assumption]{Corollary}

\newtheorem{prop}[assumption]{Proposition}
\newtheorem{remark}[assumption]{Remark}

\renewcommand{\th}{\theta}
\newcommand{\A}{\mathcal{A}}              
\newcommand{\B}{\mathcal{B}}              
\newcommand{\del}{\partial}                    
\newcommand{\DD}{\mathcal{D}}           
\newcommand{\eps}{\varepsilon}            
\newcommand{\Ga}{\Gamma}                  
\newcommand{\ga}{\gamma}                   
\renewcommand{\H}{\mathcal{H}}           
\newcommand{\half}{{\mathchoice{\thalf}{\thalf}{\shalf}{\shalf}}}
\newcommand{\<}{\langle}       
     
\newcommand{\N}{\mathbb{N}}             
\newcommand{\om}{\omega}                 
\newcommand{\ol}{\overline}                  

\newcommand{\R}{\mathbb{R}}               

\newcommand{\set}[1]{\{\,#1\,\}}               
\newcommand{\shalf}{{\scriptstyle\frac{1}{2}}} 
\renewcommand{\SS}{\mathcal{S}}        
\newcommand{\thalf}{\tfrac{1}{2}}            
\newcommand{\Afr}{\mathfrak{A}}           
\newcommand{\wh}{\widehat}                  
\newcommand{\wt}{\widetilde}                 

\def\<#1,#2>{\langle#1\,,\,#2\rangle}      

\newcommand{\be}{\begin{enumerate}}
\newcommand{\ee}{\end{enumerate}}

\newbox\ncintdbox \newbox\ncinttbox
\setbox0=\hbox{$-$} \setbox2=\hbox{$\displaystyle\int$}
\setbox\ncintdbox=\hbox{\rlap{\hbox
    to \wd2{\kern-.1em\box2\relax\hfil}}\box0\kern.1em}
\setbox0=\hbox{$\vcenter{\hrule width 4pt}$}
\setbox2=\hbox{$\textstyle\int$}
\setbox\ncinttbox=\hbox{\rlap{\hbox
    to \wd2{\kern-.14em\box2\relax\hfil}}\box0\kern.1em}

\newcommand{\sg}{\sigma}                              

\begin{document}

\thispagestyle{empty}

\begin{center}

CENTRE DE PHYSIQUE TH\'EORIQUE$\,^1$\\
CNRS--Luminy, Case 907\\
13288 Marseille Cedex 9\\
FRANCE\\

\vspace{3cm}

{\Large\textbf{Spectral action for torsion with and without boundaries}} \\
\vspace{0.5cm}
\vspace{0.5cm}

{\large  B. Iochum$^{1, 2}$, C. Levy$^{3}$ and D. Vassilevich$^{4, 5}$} \\

\vspace{1.5cm}

{\large\textbf{Abstract}}
\end{center}
We derive a commutative spectral triple and study the spectral action for a rather general geometric setting 
which includes the (skew-symmetric) torsion and the chiral bag conditions on the boundary. The spectral action 
splits into bulk and boundary parts. In the bulk, we clarify certain issues of the previous calculations, show that 
many terms in fact cancel out, and demonstrate that this cancellation is a result of the chiral symmetry of spectral 
action. On the boundary, we calculate several leading terms in the expansion of spectral action in four 
dimensions for vanishing chiral parameter $\theta$ of the boundary conditions, and show that $\theta=0$ 
is a critical point of the action in any dimension and at all orders of the expansion.

\vspace{1.5cm}



\noindent CPT-P050-2010

\noindent CPH-SYM-00
\vspace{2cm}

{\small
\noindent $^1$ UMR 6207

-- Unit\'e Mixte de Recherche du CNRS et des
Universit\'es Aix-Marseille I, Aix-Marseille II et de l'Universit\'e
du Sud Toulon-Var (Aix-Marseille Universit\'e)

-- Laboratoire affili\'e \`a la FRUMAM -- FR 2291\\
$^2$ Also at Universit\'e de Provence, iochum@cpt.univ-mrs.fr\\
$^3$ University of Copenhagen, Department of Mathematical Sciences, Denmark, levy@math.ku.dk\\  
$^4$ CMCC -- Universidade Federal do ABC, Santo Andr\'e, S.P., Brazil, dmitry.vasilevich@ufabc.edu.br\\
$^5$ Also at the Department of Theoretical Physics, St.~Petersburg University, Russia
}

\newpage


\section{Introduction}

The central object of the noncommutative geometry approach \cite{Book} is given by a spectral
triple $(\A,\H,\DD)$ consisting of an algebra $\A$, a Hilbert space $\H$ and an unbounded selfadjoint operator 
$\DD$, like the Dirac operator for manifolds.  By making use of the Dirac operator, one can define a spectral 
action\cite{CC} which becomes a natural action for a classical bosonic field theory.
Practical calculations of this action are usually done by using the heat kernel expansion, which is a standard 
instrument of quantum field theory \cite{V}. Although this approach was initially designed for noncommutative 
spaces, it appeared to be rigid enough to make predictions in the commutative case as well.
For example, being applied to Riemannian manifolds with boundary \cite{CC2}, the spectral action reproduces 
correctly the Einstein--Hilbert action together with the boundary term  \cite{HH}. It seems also natural to extend 
this approach to the Riemann--Cartan manifolds and bring into the focus a non-zero torsion.    

The spectral action  associated to a manifold endowed with a connection with torsion has been computed recently 
in \cite{HPS}, though technical tools were ready long ago \cite{Goldthorpe,Grensing,Obukhov}. The extension 
by torsion \cite{HPS} of the almost commutative spectral triple of the standard model in particle physics \cite{CCM} 
demonstrated that torsion becomes coupled to the Higgs field and, therefore is possibly observable.
  
The purpose of the present paper is to reconsider the calculations of \cite{HPS}, extend the spectral triple with 
torsion to manifolds with boundaries, and to compute the corresponding spectral action. We work mostly in four 
dimensions, although some results are valid for arbitrary even number of dimensions. As in \cite{HPS} we restrict 
ourselves to totally skew-symmetric torsion. The presence of a skew-symmetric torsion does not modify the 
geodesics and is, therefore, a rather minimalist modification of the Riemann geometry. Regarding physical 
aspects of the space-time torsion, the interested reader can consult \cite{Shapiro:2001rz} (and also 
\cite{Kleinert:2010at} for some recent developments, in particular recent limits on torsion given by two different 
experiments which place careful bounds on the four axial components of torsion \cite{KRT,HACC}).

In an expansion in the large cut-off parameter $\Lambda$ the spectral action is local, i.e. it is given as a sum of 
volume and boundary integrals. The volume part of the spectral action with torsion has been calculated in 
\cite{HPS}, but, as we show below, many terms given in that paper in fact cancel out. In particular, there is no term 
quadratic in the torsion and linear in the Riemann curvature, no quartic torsion term, etc. These cancellations are 
not accidental, but they are forced by an important symmetry of the spectral action, which is the chiral symmetry. 
The importance of chiral symmetry for the spectral action has been stressed recently \cite{Andrianov:2010nr} in a 
somewhat different context.

This prominent role of the chiral symmetry has motivated us to introduce a chiral phase also on the boundary, i.e. 
to consider the so called chiral bag boundary conditions. These are local boundary conditions depending on a real 
parameter $\theta$. They are ``mixed oblique" in the sense that in the definition of the domain of squared Dirac 
operator there is a piece which restricts solely the value  of the field at the boundary (as in the Dirichlet case), and 
a piece which contains the normal derivative at the boundary (as in the Neumann case).
This latter piece depends also on the derivatives in the directions tangential to the boundary, and this makes the 
boundary conditions ``oblique". The case $\th=0$ corresponds to the usual local mixed boundary condition.

The bag ($\th=0$) and chiral bag ($\th\ne 0$) boundary conditions appeared first in theoretical physics, in the 
models of hadrons (see \cite{Hasenfratz:1977dt} for a review). Properties of the determinants of the Dirac operator 
for chiral bag conditions and their relations to chiral symmetry breaking were studied in \cite{Hrasko:1983sj,Wipf}.
A more detailed mathematical investigation of the spectral properties of the Dirac operator for these conditions 
was performed in \cite{EGK,EK,GK}. 
 
The goal of this paper is to compute the spectral action for spectral triples on compact 4-dimensional manifolds 
with boundary already considered in \cite{CC2,CC3} but in presence of a totally skew-symmetric torsion $T$. In 
order to do so, we exhibit simple spectral triples $(\A,\DD,\H)$ with vanishing tadpoles (Theorem \ref{triple}) in the 
case of a chiral bag boundary condition. Essentially, the construction follows the same arguments of \cite{Boundary}, 
with one difference: the boundary operator $\Pi_-$, related to the chirality $\chi_\th$, is not selfadjoint. In respect to 
this, we extend \cite[Theorem 4.5]{Boundary} to non-selfadjoint boundary operators in Theorem 
\ref{thmboundary}.  
 
We generalize \cite{HPS} for chiral mixed boundary condition (in the case $\th=0$) by computing the spectral 
action in dimension 4 in section \ref{secaction}. This restriction to $\th=0$ is justified by the fact that this is a critical 
point for the spectral action, a result obtained in section 5 via the Index Theorem.

\section{Notations}

Let $\wt M$ be a smooth compact Riemannian (with the metric $g$) manifold without boundary of even dimension 
$n=2m$. We denote $\nabla^{LC}$ the Levi-Civita connection associated to $g$. We fix a linear connection 
$\nabla^{g}$ on $T\wt M$ which is $g$-compatible and with the same geodesics as $\nabla^{LC}$. Equivalently, 
$\nabla^{g}$ is a linear connection with a totally skew-symmetric torsion $8T$ (the chosen coefficient 8 is justified 
below).

It is known in this case that we have $\nabla^g_XY = \nabla^{LC}_XY + 4 T(X,Y,-)$, or in other words, for any 
vector fields $X,Y,Z$, 
\begin{equation}
\label{torsionrelation}
g(\nabla^g_XY - \nabla^{LC}_XY,Z) = 4 T(X,Y,Z). 
\end{equation}

Let $M$ be a submanifold of $\wt M$ of dimension $n$ such that its topological closure $\ol M$  is a compact 
manifold with smooth boundary $\partial M=\ol M \backslash M$. This implies that $\partial M$ is a smooth 
compact submanifold of $\wt M$ without boundary of dimension $n-1$. 
Remark that, for a given manifold $M$ with boundary $\partial M$, there exists a manifold $\wt M$ with previous 
property.

We also assume that $\wt M$ is endowed with a complex vector bundle $\wt V$ of dimension $2^m$ and with a 
smooth map $\ga : T \wt M \to \End(\wt V)$ giving a structure of Clifford module to $\wt V$. This means that for any 
$x,y\in T\wt M$, $\set{\ga(x),\ga(y)}=2g(x,y)$. 

Let us remark immediately that the convention chosen here $(+2g(x,y))$ for the Clifford anticommutation relations 
is different from the one $(-2g(x,y))$ used for instance in \cite{BGKS,BG2,Gilkey,Gilkey2,GK}. One can go from 
one convention to another with a multiplication by $i$. More precisely, $\ga':=i\ga$ satisfies 
$\set{\ga'(x),\ga'(y)}=-2g(x,y)$. 

We also fix on $\wt V$ a Hermitian inner product $(\cdot,\cdot)$ such that $\ga(x)^*=\ga(x)$ for any $x\in T\wt M$. 
Note that for a given $\ga$, such Hermitian inner product always exists. In the following, we shall say that a given 
connection $\nabla^{\wt V}$ on $\wt V$ is \emph{unitary} if for any $x\in T\wt M$ and $v,w\in \wt V$
$$
x(v,w) = (\nabla_x^{\wt V} v ,w) + (v,\nabla_x^{\wt V} w)
$$
and \emph{compatible with} $(\nabla^{T\wt M},\ga)$, where $\nabla^{T\wt M}$ is a given $g$-compatible 
connection on $T\wt M$ with totally skew-symmetric torsion, if for any $x,y\in T\wt M$ and $v\in \wt V$, 
$$
\nabla_x^{\wt V} \big(\ga(y) v \big) = \ga \big(\nabla_x^{T\wt M}(y) \big) v + \ga(y)\nabla^{\wt V}_x v\, .
$$
Our conventions are the following (see \cite{V}): $(e):=\set{e_1,\ldots,e_n}$ is a local orthonormal frame of the 
tangent space where $e_n$ is the inward pointing unit vector field. We assume that Roman indices $a,b,c$ range 
from 1 to $n-1$ and index an orthonormal frame for the tangent bundle of $\partial M$. Local coordinates charts 
are denoted $(x_\mu)$, and $(\del_\mu)$ is the associated local coordinate frame. We shall use the shorthands: 
$\ga_i:=\ga(e_i)$, $\ga_\mu:=\ga(\del_\mu)$ and $\ga^\mu:=g^{\mu\nu}\ga_\nu$. The coordinates of $e_i$ in the 
frame $(\del_\mu)$ are denoted $(e_i^\mu)$ and the inverse matrix is denoted $(e^i_\mu)$.

By the proof of \cite[Lemma 1.1.7]{Gilkey2}, if we fix a given partition of unity on $\wt M$ and associated local 
frames for the bundle $\wt V$, we can construct a connection on $\wt V$ which is unitary and compatible with 
$(\nabla^{LC},\ga)$. The idea is to patch together the connections locally defined by the matrices
\begin{equation}
\label{omegasi}
\om_i^{s} := -\tfrac{1}{4} \Ga_{ijk}\ga_j^s\ga_k^s
\end{equation}
where $s=(s_p)$ is a local frame for $\wt V$, $(e)$ is a local orthonormal frame with $e_i \ga_j^s=0$, $\Ga_{ijk}$ 
are the Christoffel coefficients of the first kind of the Levi-Civita connection in $(e)$, and $\ga_j^s$ is the matrix of 
$\ga_j$ in the frame $s$. Note that the minus sign in \eqref{omegasi}, not present in \cite{Gilkey2}, comes from our 
different definition of $\ga$.

Actually, we can see from the proof that the same construction using $\nabla^{g}$ instead of $\nabla^{LC}$ is still 
valid, since the fact that $\nabla^{LC}$ is torsion-free is not used. The crucial point in the proof is the relation 
$\Ga_{ijk}=-\Ga_{ikj}$, which is true for any $g$-compatible connection on $T\wt M$.

As a consequence, we can apply \cite[Lemma 1.1.7]{Gilkey2} to define a ``spin'' connection $\nabla^{(T)}$ on 
$\wt V$ which is unitary and compatible with $(\nabla^g,\ga)$. The same construction using $\nabla^{LC}$ yield to 
the connection $\nabla^{(0)}$ on $\wt V$ which is unitary and compatible with $(\nabla^{LC},\ga)$.
Locally, since $T$ is skew-symmetric, \eqref{torsionrelation} and \eqref{omegasi} give
$\nabla_j^{(T)}=\nabla^{(0)}_j- T_{jkl}\,\ga_k \ga_l$, where $T_{jkl}:=T(e_j,e_k,e_l)$.
We define the exterior product $\wedge$ so that $T=3! \sum_{j<k<l} T_{jkl}\,\th_j \wedge \th_k \wedge \th_l$, 
where $(\th_j)$ is the dual coframe of $(e_j)$. The Clifford action of $T$ (still denoted $T$) is thus 
\begin{align}
T=6\sum_{j<k<l}T_{jkl} \,(i\ga_j)(i\ga_k)(i\ga_l)=-i\sum_{jkl} T_{jkl}\, \ga_j\ga_k\ga_l.
\label{defT}
\end{align}
The Dirac operators respectively associated to $\nabla^{(T)}$ and $\nabla^{(0)}$ are
\begin{align*}
\DD^{(T)} :=i\sum_{j=1}^n \ga_j \nabla^{(T)}_{{j}}\, , \qquad \DD^{(0)} :=i\sum_{j=1}^n \ga_j \nabla^{(0)}_{{j}},
\end{align*}
so we deduce from the preceding equations that 
$$
\DD^{(T)} = \DD^{(0)} + T
$$
and this relation hence justifies the factor 8 in the definition of the torsion of $\nabla^g$. 
 
The Dirac operators $\DD^{(T)}$ and $\DD^{(0)}$ are formally selfadjoint, as 
unbounded operators on $\H:=L^2(\wt M,\wt V)$, at least when $\wt M$ has no boundary \cite{FS}. Of course, since 
we want to deal with a spectral triple $(\A,\DD^{(T)},\H)$, we need a selfadjoint $\DD^{(T)}$.

We denote $\Ga^{k}_{ij}$ the Christoffel coefficients of $\nabla^{LC}$. If $R_{ijkl}$ is the component of the 
Riemann curvature tensor for $\nabla^{LC}$, the Ricci tensor is $R_{jk}:=R_{ijki}$ with scalar curvature $R=R_{ii}$ 
(so $R=+12$ for the 4-sphere).

Moreover, we note $L_{ab}:=\Gamma_{ab}^n$ the second fundamental form or extrinsic curvature and 
$L:=L_{aa}$. 

In coordinates, we get $(\nabla^{LC}_\mu X)_\nu=\partial_\mu X_\nu+\Gamma_{\mu\rho}^\nu X_\rho$. At the 
Clifford module level, we get, using previous notations, 
$(\nabla_\mu^{(0)} v)_{p}=\partial_\mu v_p + (\om_\mu^s(v))_p$, where $(\om_\mu^s)$ is the connection 1-form 
associated to $\nabla^{(0)}$ and the frame $s=(s_p)$. A computation shows that 
$\om^s_\mu=\tfrac{1}{4}\sg_\mu^{\nu\rho}\ga^s_\nu \ga^s_\rho=\tfrac{1}{8}\sg_\mu^{\nu\rho}[\ga^s_\nu,\ga^s_\rho]$ 
where the spin connection $(\sg_{\mu}^{\nu\rho})$ is locally defined by $-e_\mu^j e_k^{\nu} e_l^\rho \Ga_{jkl}$.

\section{The spectral triple for chiral bag boundary condition}

As boundary conditions for the operator $\DD^{(T)}$ and its square, we prefer to choose local ones which 
guarantee its selfadjointness on $M$, as the chiral bag boundary condition which is based on the existence of the  
usual chirality $\ga_{n+1}$ (which exists on even dimensional manifolds) and a real parameter $\theta$. 

It has been shown \cite{BGKS} that such chiral bag boundary conditions yield a strongly elliptic boundary value 
problem for the Dirac operator and its square. Moreover, the associated heat-kernel asymptotics has been 
investigated on the Euclidean ball in \cite{EK} and in a more general setting in \cite{EGK}. Some stability 
properties of heat-kernel coefficients with respect to parameter $\th$ have been established in \cite{GK}. 

We give in this section the construction of a spectral triple based on a Dirac operator with a chiral bag boundary 
condition. 

Recall from \eqref{defT} that the 3-form $T$ acts on the Clifford module as $-i T_{jkl} \,\ga_{j}\ga_k\ga_{l}$, thus 
$T$ is formally selfadjoint as an operator (acting by Clifford multiplication) on $L^2(\wt M,\wt V)$.
Recall also that $\DD^{(T)}=\DD^{(0)}+T$ where $\DD^{(0)}$ is the usual Dirac operator based on the spin 
connection compatible with $(\nabla^{LC},\ga)$. 

The chirality matrix is defined by 
$$
\ga_{n+1}:=(-i)^{n/2}\,\ga_1\cdots\ga_n\,,
$$ 
so $\ga_{n+1}^*=\ga_{n+1}$, $\ga_{n+1}^2=\text{Id}_V$ and $\ga_i\ga_{n+1}=-\ga_{n+1}\ga_{i}$. 

\noindent Note that $\set{\ga_{n+1},\DD^{(T)}}=\set{\ga_{n+1},\DD^{(0)}}=\set{\ga_{n+1},T}=0$ and if $n=4$, 
$\ga_5=-\ga_1\cdots\ga_4$. 

We also define
$$
\chi_\th:= -i e^{\th \ga_{n+1}}\ga_{n+1} \ga_n\,.
$$
Since $\ga_n$ anti-commutes with $\ga_{n+1}$, $\chi_\th^2=\text{Id}_V$, so 
$\Pi_{\pm}:=\tfrac{1}{2}(\text{Id}_V \pm \chi_\th)$ are two non-selfadjoint idempotents. A direct computation 
shows that ${\chi_\th}^* = \chi_{-\th}$ and
\begin{equation}
\label{eqgammachi}
\chi_{-\th}\, \ga_n + \ga_n\, \chi_\th = 0, \qquad \ga_n + \chi_{-\th} \,\ga_n\, \chi_\th =0\, .
\end{equation}

The sub-bundle of $\wt V$ on $\ol M$ (resp. $\del M$) is denoted $\ol V$ (resp. $V$) and $H^s(\wt V)$, 
$H^s(\ol V)$ are the Sobolev spaces of order $s\in \R$ respectively on $\wt M$ with bundle $\wt V$ and $\ol M$ 
with bundle $\ol V$. Recall that by definition 
$$
H^s(\ol V) := H^s(\ol M,\ol V):= r_+ \big(H^s(\wt V)\big)\, 
$$
where $r_+$ is the restriction on $M$.
We define
$$
C^\infty(\ol M, \ol V):= r_+\,\big( C^\infty(\wt M,\wt V) \big) \, , \qquad
C^\infty(\ol M):=r_+ \big(C^\infty(\wt M)\big) \, .
$$
The extension by zero operator $e_+$ is a linear continuous operator from $H^s(\ol V)$ into $H^s(\wt V)$ for 
any $s\in ]-\half,\half[$ such that $e_+(u)=u$ on $M$ and $e_+(u)(x)=0$ for any 
$u \in C^\infty(\ol M,  \ol V)$ and $x\in \wt M\backslash M$. 

When $P$ is a pseudodifferential operator on $(\wt M,\wt V)$, its truncation to $M$ is given by
$$ 
P_+ := r_+ \, P \, e_+\, .
$$
Let $\DD^{(T)}_{\B_{\chi_\th}}$ be the realization of $\DD^{(T)}$ on $\B_{\chi_\th}$, that is the operator acting as 
$\DD^{(T)}_+$ on the space of sections $\psi$ in $H^1(\ol V)$ satisfying the boundary condition 
$\B_{\chi_{\th}} \psi=0$ with 
$$
\B_{\chi_{\th}} \psi:=\Pi_-\, r(\psi)\,
$$
where we denoted $r$ the restriction operation on $\del M$ (denoted $\ga_0$ in \cite{Grubb}). 

As in \cite{Boundary}, define $ \A_{\DD^{(T)}_{\B_{\chi_\th}}}$ as the $*$-algebra of smooth functions 
$a \in C^\infty(\ol M)$ such that $a$ and $a^*$ apply $\H^\infty_{{\DD^{(T)}_{\B_{\chi_\th}}}}$ into itself where  
$$
\H^\infty_{{\DD^{(T)}_{\B_{\chi_\th}}}}:=\bigcap_{k\geq 1} \, \text{Dom } \DD^{(T)}_{\B_{\chi_\th}}.
$$
These functions act as multiple of the identity on the Hilbert space $L^2(\ol V)$.

We refer to \cite{Grubb} for the definition of an elliptic pseudodifferential boundary system and recall the Green 
formula \cite[1.3.2 Proposition]{Grubb}: if $P$ is a pseudodifferential operator of order $k\in \N$, then for any 
$u,v\in C^\infty(\ol M,\ol V)$,  
\begin{align*}
(P_+ u ,v )_M - (u,(P^*)_+ v)_M = (\Afr_P\, \rho u , \rho v)_{\del M}
\end{align*}
where $\rho=\set{r_0,\cdots,r_{k-1}}$ is the Cauchy boundary operator defined by 
$r_j u=(-i\partial_n)^j\, u_{\vert \del M}$ with $r_0:=r$ (here, $\partial_n$ is the interior normal derivative) and 
$\Afr_P$ is the Green matrix associated to $P$. When $k=1$, $\Afr_P$ is an endomorphism on the boundary 
$\del M$. In particular, $\Afr_{\DD^{(0)}}=-i\ga_n$.

\bigskip

It is known that $\DD^{(T)}_{\B_{\chi_\th}}$ is selfadjoint \cite[Theorem 2.1]{BGKS}, but we give here a proof 
based on another approach more appropriate to noncommutative geometry. We begin with an extension of 
\cite[Theorem 4.5]{Boundary} in which the boundary endomorphism $\Pi$ is no more supposed to be selfadjoint:
 
\begin{theorem}
\label{thmboundary}
Let $P\in \Psi^1(\wt V)$ be a symmetric pseudodifferential operator of order one on $\wt M$ satisfying the 
transmission property (see \cite[1.2]{Grubb}).

Let $\Pi \in C^\infty\big(\del M,L( V)\big)$ be an idempotent endomorphism on the boundary such that the system 
$\set{P_+,\B:=\Pi \, r}$ is an elliptic pseudodifferential boundary operator.  Then 

(i) $P_\B$ is selfadjoint if and only if
\begin{align}
\label{Afr}
(1-\Pi^*) \,\Afr_P\,(1-\Pi)=0 \quad \text{and} \quad \Pi\, \Afr_P^{-1}\, \Pi^* =0 \, .
\end{align}

(ii) When $P_\B$ is selfadjoint, $\big(C^\infty(\ol M), L^2(\ol V), P_\B\big)$ is a spectral triple of dimension dim(M).

(iii) When $P$ is a differential operator such that $P^2$ has a scalar principal symbol and $P_\B$ is 
selfadjoint, the spectral triple $\big(\A_{P_\B}, L^2(\ol V), P_\B\big)$ is regular.

(iv) Under the hypothesis of $(iii)$, $\A_{P_\B}$ is the largest algebra $\A$ in $C^\infty(\ol M)$ such that the triple
$\big(\A, L^2(\ol V), P_\B\big)$ is regular.
\end{theorem}

\begin{proof}
$(i)$ For any given endomorphism $R$, we shall denote $\wh  R$ the surjective morphism defined as the 
operator acting as $R$ from the domain of $R$ into the image of $R$. 

Since $P$ is a pseudodifferential operator of order 1 and $\wh \Pi$ is a surjective morphism, we can apply 
\cite[1.6.11 Theorem]{Grubb} with the following choices: the $S$ of \cite{Grubb} is $\wh \Pi$ and $S':=\wh {1-\Pi}$. 
Note that $\Pi$ is here not surjective: it is an 
endomorphism only surjective on $V^+ := \Pi(V)$ with kernel $V^-:= (1-\Pi)(V)$, so $V$ is
the direct (not necessarily orthogonal) sum of $V^+$ and $V^-$. In the notation of \cite{Grubb}, we take $B=P_\B$, 
$G=K=G'=\wt G=T'=0$ and $\rho=r$. Remark that $\Afr_P^*=-\Afr_P$ since $P=P^*$ (viewed as defined on 
$H^1(\wt M, \wt V)$). Thus we have (still using the notations of \cite{Grubb}), $T=\B$ and 
$\wt T = {C'}^* \Afr_P^* \, r=-{C'}^*\Afr_P\, r$ (the matrix $I^\times$ is the number 1 here). 

According to \cite[1.6.11 Theorem]{Grubb}, since $\set{P_+,\B}$ is elliptic, $P_\B$ is selfadjoint if there is a 
homeomorphism $\Psi$ from $H^s(V^+)$ onto $H^s(V^-)$, such that
\begin{equation}
-C'^*\, \Afr_P \,r = \Psi \, \wh  \Pi \, r\, ,\label{eqadjoint}
\end{equation}  
with $C'$ satisfying $\wh {(1-\Pi)} C'=\text{Id}_{V^-}$ and $ C' \wh {(1-\Pi)}=1-\Pi$. In other words, $C'$ is the 
injection from $V^-$ into $V$. 

By \cite[(1.6.52)]{Grubb}, when this is the case, $\Psi$ has the form $\Psi=C'^*\,\Afr_P^*\,C$ with 
$\wh{\Pi}\,C=\text{Id}_{V^+}$ and $C \,\wh{\Pi}=\Pi$. Note that $\Afr_P$ is invertible as a consequence of the 
ellipticity of $P$. 

The following computation shows that $C'^*=P_- (1-\Pi^*)(=P_-)$, where $P_-$ is the orthogonal projection from 
$V$ onto $V^-$: for any $u\in V$ and $v\in V^-$, 
$$
(C'^* u, v)_{V^-} = (u,C'v)_{V}=(u,v)_{V}= ((1-\Pi^*)u,v)_{V}=(P_-(1-\Pi^*)u,v)_{V^-}\, .
$$
Now, suppose that $(1-\Pi^*)\, \Afr_P\,(1-\Pi)=0$ and $\Pi\, \Afr_P^{-1}\, \Pi^*=0$. We define
$\Psi=-C'^*\,\Afr_P \,C$ which is a homeomorphism from $H^s(V^+)$ onto $H^s(V^-)$. Indeed, 
if we set $\Psi^{-1} := - \wh {\Pi}\, \Afr_P^{-1} (\wh {1-\Pi})^*$, we get
\begin{align*}
\Psi \circ \Psi^{-1} &= C'^* \,\Afr_P \,C \,\wh \Pi \,\Afr_P^{-1}\, (\wh {1-\Pi})^* 
=  P_-{(1-\Pi^*)}\,\Afr_P \,\Pi \,\Afr_P^{-1} \,(\wh {1-\Pi})^* \\
&=  P_- {(1-\Pi^*)}\, \Afr_P \,\big(\Pi +(1-\Pi)\big) \,\Afr_P^{-1} \,(\wh {1-\Pi})^* = C'^* \, (\wh {1-\Pi})^* = \text{Id}_{V^-}\,,
\end{align*}
and 
\begin{align*}
\Psi^{-1} \circ \Psi &= \wh \Pi \,\Afr_P^{-1} (\wh{1-\Pi})^* \,C'^* \,\Afr_P\, C 
=  \wh \Pi \,\Afr_P^{-1} (1-\Pi^*)\,\Afr_P\, C  \\
&=  \wh \Pi \,\Afr_P^{-1} (1-\Pi^*+\Pi^*)\,\Afr_P\, C  = \wh \Pi\, C =  \text{Id}_{V^+}\,.
\end{align*}
Moreover,  
$$   
\Psi \, \wh \Pi = -C'^* \, \Afr_P \, \Pi = -P_- (1-\Pi^*) \,\Afr_P\, \big(\Pi +(1-\Pi)\big) = -C'^* \,\Afr_P\,.
$$
As a consequence, \eqref{eqadjoint} is satisfied and the if part of the assertion follows.

Conversely, suppose that $P_\B$ is selfadjoint. Using Green's formula, $\big(\Afr_P\, r (u) , r (v)\big)_{\del M}=0$ for 
any $u,v\in \Dom P_\B$. Since $r \, : \, H^1(\ol V) \rightarrow H^{1/2}(V)$ is surjective, 
$\big(\Afr_P\, (1-\Pi)\psi , (1-\Pi) \phi \big)_{\del M}=0$ for any $\psi,\phi \in H^{1/2}(V)$ and thus 
$(1-\Pi^*)\, \Afr_P \,(1-\Pi)=0$. Again, from 
\cite[Theorem 1.6.11]{Grubb} we get that $\Psi:= C'^* \,\Afr_P^* \, C$ is a homeomorphism from $H^s(V^+)$ onto 
$H^s(V^-)$ and we check as before that $\Psi^{-1}:= - \wh {\Pi}\, \Afr_P^{-1} (\wh {1-\Pi})^*$ is a right-inverse of 
$\Psi$, and thus, is the inverse of $\Psi$. 

\noindent The equation $\Psi^{-1}\circ \Psi = \Id_{V^+}$ yields 
$\wh \Pi\, \Afr_P^{-1} \,\Pi^* \,\Afr_P \,C=0$, which gives $\wh \Pi\,\Afr_P^{-1} \,\Pi^*\, \Afr_P \,\Pi=0$. Thus, 
\begin{align*}
\wh \Pi\, \Afr_P^{-1} \,\Pi^* \,\Afr_P = \wh \Pi\, \Afr_P^{-1} \,\Pi^* \,\Afr_P \,(1-\Pi) = 
\wh \Pi \,\Afr_P^{-1} \,\big(\Pi^* +(1-\Pi^*) \big)\,\Afr_P \,(1-\Pi)= \wh \Pi \,(1-\Pi) =0
\end{align*}
so $\Pi\, \Afr_P^{-1}\, \Pi^*=0$. 

$(ii,iii,iv)$ The corresponding proofs of \cite[Theorem 4.5]{Boundary} can be applied directly. 
\end{proof}

\begin{corollary}
\label{D=D*}
$\DD^{(T)}_{\B_{\chi_\th}}$ is a selfadjoint operator.
\end{corollary}

\begin{proof}
Since it is a Dirac operator, $\DD^{(0)}$ is an elliptic symmetric differential operator of order one such 
that the principal symbol of its square is scalar, and it is selfadjoint with domain $H^1(\wt V)$. 

Since the principal symbol is unchanged by the perturbation $\DD^{(0)}\to \DD^{(T)}=\DD^{(0)}+T$, and $T$ is 
selfadjoint, $\DD^{(T)}$ is also an elliptic symmetric differential operator of order one such that the principal 
symbol of its square is scalar, and it is selfadjoint with domain $H^1(\wt V)$. 

Moreover, the Green operator $\Afr_{\DD^{(T)}}$ of $\DD^{(T)}$ coincides with 
$\Afr_{\DD^{(0)}}= -i\ga_n=-\Afr^{-1}_{\DD^{(0)}}$ and we get by \eqref{eqgammachi}
\begin{align*}
&(1-\Pi_-^*)\ga_n (1-\Pi_-) = \Pi_+ ^*\ga_n \Pi_+ =  \tfrac{1}{4}(\ga_n+\chi_{-\th}\,\ga_n +\ga_n \,\chi_\th 
+ \chi_{-\th}\, \ga_n \,\chi_\th)=0 ,\\ 
&\Pi_-\ga_n \Pi_-^* =  \tfrac{1}{4}(\ga_n-\chi_\th \, \ga_n -\ga_n \,\chi_{-\th} + \chi_\th \,\ga_n \,\chi_{-\th})=0 .
\end{align*}
Thus, \eqref{Afr} is satisfied and the claim follows from previous theorem $(i)$.
\end{proof}
If $\wt M$ (with $n$ even) has a spin structure with Dirac operator $\DD^{(0)}$, and $J$ denotes the ordinary 
charge conjugation on the spin manifold $\wt M$, then $\set{J,\ga_i}=0$. So $[J,T]=[J,\DD^{(0)}]=0$. Moreover, 
$J \ga_{n+1} = \eps' \ga_{n+1} J$, and thus $J\chi_\th = \eps' \chi_{\eps'\th} J$, where $\eps'=-1$ if $n/2$ is odd and 
$\eps'=1$ if $n/2$ is even. 

If we define $J':= J$ if $n/2$ is even and $J':=J\ga_{n+1}$ if $n/2$ is odd, then
$J'$ is an antilinear isometry satisfying $ \DD^{(T)}  J' = (-1)^{n/2} J' \DD^{(T)}$, $ J' b  J'^{-1} = b^*$ for all 
$b\in C^\infty(\wt M)$, so is a conjugation operator as defined in \cite[Definition 5.2]{Boundary}. Moreover 
$J' \chi_\th =  \chi_{(-1)^{n/2}\th} J'$, so if $\th=0$ when $n/2$ is odd, 
$\DD^{(T)}_{\B_{\chi_\th}} J_+ = (-1)^{n/2} J_+ \DD^{(T)}_{\B_{\chi_\th}}$.

Thus, we get a spectral triple for a Dirac operator on an even dimensional manifold endowed with a chiral bag 
boundary condition and a totally skew-symmetric torsion:

\begin{theorem} 
\label{triple}
For any $\theta \in \R$, $\big( \A_{\DD^{(T)}_{\B_{\chi_\th}}}\, ,\,L^2(\ol M,\ol V)\,,\,\DD^{(T)}_{\B_{\chi_\th}} \big)$ is 
a spectral triple which has a simple dimension spectrum and is regular. Moreover, if $\wt M$ is a spin manifold with 
Dirac operator $\DD^{(0)}$, and if $\th=0$ when $n/2$ is odd, this triple has no tadpoles.
\end{theorem}
   
\begin{proof} 
The proof follows from the same arguments of \cite[Theorem 4.8, Theorem 5.8]{Boundary}, using Theorem 
\ref{thmboundary} instead of \cite[Theorem 4.5]{Boundary} and $J'_+$ being a conjugation operator for the spectral 
triple $\big( \A_{\DD^{(T)}_{\B_{\chi_\th}}}\, ,\,L^2(\ol M,\ol V)\,,\,\DD^{(T)}_{\B_{\chi_\th}} \big)$ in the sense of 
\cite[Definition 5.2]{Boundary}.
\end{proof}

\section{The spectral action}

\subsection{Mixed boundary condition: $\theta=0$}

We suppose from now on that the dimension is $n=4$.

We can rewrite $T$ as a linear combination of $-i\ga_j\ga_{n+1}$ matrices. More precisely, we have 
$T=-i\ga_j\ga_{n+1} T_j$ where $T_j := -  \eps_{jpkl} \,T_{pkl} \in C^\infty(\wt M,\R)$.
In coordinates, $T_\mu = e^{j}_\mu \,T_j$. In this respect, the Dirac operator reads 
\begin{align}
\left\{\begin{array}{l}
\DD^{(T)}=\DD^{(0)} + T,  \\   
\DD^{(0)}=i\gamma^\mu \left( \partial_\mu +\tfrac 18 [\gamma_\nu ,\gamma_\rho ]
\sigma_\mu^{\nu\rho}\right), \qquad T= i\gamma^\mu \left(-\gamma_{n+1} T_\mu \right),
\end{array}\right.
\label{Dirop}
\end{align}
where $(\sg_\mu^{\nu\rho})$ is the spin connection. Remark that $A_\mu=0$ and $A_\mu^5=iT_\mu$ in the 
notation of \cite[(3.27)]{V}). Note also that $A_\mu^5$ does not carry any spinor indices, which is a particularity of 
the chosen dimension $n=4$.

The square of $\DD^{(T)}$ produces a Laplace type operator on $C^\infty(\wt V)$ and by 
\cite[Lemma 4.8.1]{Gilkey} there is a unique connection $\nabla^{{\DD}^{(T)}}$ on $\wt V$ and a unique 
endomorphism $E \in C^\infty\big(\text{End}(\wt V)\big)$ so that 
\begin{align}
{\DD^{(T)}}^2= - \big(g_{\mu \nu} \nabla^{{\DD}^{(T)}}_\mu \nabla^{{\DD}^{(T)}}_\nu +E \big)
\end{align}
where
\begin{align}
&\nabla^{{\DD}^{(T)}}_\mu =\partial_\mu + \omega_\mu,\\
&\omega_\mu :=-\tfrac 12 [\gamma_\mu,\gamma_\nu] T_\nu \gamma_5
+\tfrac 18 [\gamma_\rho ,\gamma_\sigma ]\sigma_\mu^{\rho\sigma},\label{om}\\
&E:=2T^2-\tfrac 14 R -\gamma_5 \nabla_\mu T_\mu .\label{E}
\end{align}
We will use the shorthand notations
\begin{align*}
&T^2:= T_\mu T_\mu, \\
&\gamma T:=\gamma_\mu T_\mu, \\
&T_{\mu \nu}:=\nabla_\mu T_\nu-\nabla_\nu T_\mu.
\end{align*}
The field strength of the connection $\omega$ is
\begin{align*}
\Omega_{\mu\nu}=\gamma^\rho (\nabla_\mu T_\rho \gamma_\nu 
- \nabla_\nu T_\rho \gamma_\mu)\gamma_5 -T_{\mu\nu}\gamma_5 
- \tfrac 14 \gamma^\rho \gamma^\sigma R_{\rho\sigma\mu\nu}
+ (\gamma T) \gamma_\mu (\gamma T) \gamma_\nu 
- (\gamma T) \gamma_\nu (\gamma T) \gamma_\mu.
\end{align*}
We follow the boundary conditions introduced in \cite{BG2} and which have been considered in 
\cite{BGKV, CC2,Tadpole,Boundary,V}: 
choose 
\begin{align}
\chi:= \chi_0=i \ga_1\ga_2\ga_3 \label{chi}=-i\ga_5\ga_4\, .
\end{align}
Thus $\chi=\chi^*$, $\{\chi,\ga_n\}=0$ and $[\chi,\ga_a]=0$ for $a=1,2,3$. In particular, $\ga_5=i\chi\ga_4$ is the 
natural chirality of the even dimensional manifold $M$.

As above, let $\Pi_\pm:=\half(\Id_{V} \pm \chi)$ be the projections on the $\pm1$ eigenvalues of $\chi$ and 
$V_\pm:=\Pi_\pm V$ be the sub-bundles of $V$ and fix an auxiliary endomorphism $S$ on $V$. 

\noindent We define $\DD^{(T)}_\B$ as the realization of $\DD^{(T)}$ with the boundary condition 
$\B_\chi \psi=0$ where $\B_\chi$ is defined on $C^\infty(\ol V)$ by
$$
\B_\chi \psi:=\Pi_- \psi_{|\del M}\,, \quad \psi \in C^\infty(\ol V).
$$
The mixed boundary conditions $\B=\B(\chi,S)$ are defined on ${\DD^{(T)}_\B}^2$ by $\B \psi=0$ where 
\begin{align}
\B \psi := \Pi_- (\DD^{(T)}_+  \psi)_{ |\del M} \oplus
\Pi_{-}\psi_{|\del M}\,, \quad \psi \in C^\infty(\ol V). \label{BC1}
\end{align}
As in \cite{BG2}, ${\DD^{(T)}}^2$ could also have considered directly with domain given by $\B \psi=0$ with
\begin{align}
\B \psi := \Pi_+ (\nabla_{n}^{\DD^{(T)}} +S)\Pi_+ \psi_{ |\del M} \oplus
\Pi_{-}\psi_{|\del M}\,, \quad \psi \in C^\infty(\ol V). \label{BC2}
\end{align}
For the choice of $\chi =-i\gamma_5 \gamma_n$, these last two boundary conditions are equivalent 
\cite[Lemma 7]{BG2} if 
$S=\tfrac{1}{2}\Pi_+ \big(-[i\ga_n,T]-L_{aa}\chi \big) \Pi_+$, thus using $\set{\ga_{5}, \chi}=0$,
\begin{align}
S=-\tfrac{1}{2}  L_{aa}\Pi_+. \label{S}
\end{align}

These boundary conditions generalizes Dirichlet ($\Pi_-=\Id_V$, so $\chi=-{\text{id}_V}$) and Neumann--Robin 
($\Pi_+=\Id_V$, so $\chi={\text{id}_V}$) conditions.

While Corollary \ref{D=D*} is the real motivation for the above choice of boundary conditions, it is nevertheless  
worthwhile to recall as in \cite{CC2,CC3} that signs and ratios in the spectral action given by that choice, are 
identical to the Euclidean action used in gravitation \cite{HH} for a zero torsion.

\subsection{Computation of the spectral action for $\theta=0$}
\label{secaction}

The Chamseddine--Connes spectral action is 
\begin{align}
\label{action}
\SS(\DD^{(T)}_{\B_\chi},\Phi,\Lambda):=\Tr \big( \Phi({(\DD^{(T)}_{\B_\chi}})^2/\Lambda^2) \big)
\end{align}
where $\Phi$ is any positive even function viewed as a cut-off with $\Lambda\in \R$.

As detailed in \cite{ConnesMarcolli}, the spectral action \eqref{action} of the above spectral triple (Theorem 
\ref{triple}, with $\th=0$) is related to the asymptotic expansion
\begin{align}
{\rm Tr} \big( \exp(-t \, (\DD^{(T)}_{\B_\chi})^2 \big) \, \underset{t \,\downarrow \,0}{\sim}  \,\sum_{k\in \N} t^{(k-4)/2} \,
{a}_k \big(1,\DD^{(T)},\chi \big)
\label{hke}
\end{align}
via
\begin{align}
\SS(\DD^{(T)}_{\B_\chi},\Phi,\Lambda)=\Lambda^4 \Phi_4 a_0+\Lambda^3 \Phi_3 a_1 + \Lambda^2 \Phi_2 a_2 
+\Lambda \Phi_1 a_3 + \Phi(0) a_4 + \mathcal{O}(\Lambda^{-1}) 
\label{action1}
\end{align}
with $\Phi_k:=\tfrac{1}{\Ga(k/2)}\int_0^\infty \Phi(s)s^{k/2-1}ds$. (Note that the decomposition \eqref{hke} is not the 
same as in \cite[Theorem 1.145]{ConnesMarcolli} or \cite[(1)]{Boundary} so the coefficients $\Phi_k$ are different.)

So, using \cite{Gilkey2,Kirsten,V} with the decomposition of $a_k$ in $T=0\,$-part and remainder, we get
\begin{align}
& a_0^{(T)}=a_0^{(0)}= \tfrac{1}{4\pi^2} \text{Vol} (M),\\
& a_1^{(T)}=a_1^{(0)}=0,\\
& a_2^{(T)}=a_2^{(0)}+ \tfrac 1{6(4\pi)^2}  \,\int_M 48T^2, \label{a2} 
\\
& a_2^{(0)}:=-\tfrac 1{6(4\pi)^2} \big( \,\int_M 2R +\int_{\partial M} 4L \, \big).\nonumber
\end{align}
(All integrations are with respect of the volume form on $M$ and of the induced volume form on $\partial M$.) 
Using ``;" for the tangential covariant differentiation on $\widetilde{M}$ and ``:" for the same on the boundary, the  
computation of
\begin{align*}
\chi_{:a}=i T^c [\gamma_a ,\gamma_c] \gamma_n +i \gamma_5 L_{ab}\gamma_b\, ,
\end{align*}
yields
\begin{align}
\left\{\begin{array}{l}
a_3^{(T)}=a_3^{(0)}-\tfrac 1{96(4\pi)^{3/2}}\int_{\partial M}  96T_aT_a \,,\label{a3} \\
a_3^{(0)}:=\tfrac 1{96(4\pi)^{3/2}}\int_{\partial M} 3L^2-6L_{ab}L_{ab} \,. 
\end{array}\right.
\end{align} 

For the next coefficient $a_4$, we have two contributions \cite{BG4,V94} (where a common pre-factor of 
$\frac{4}{360(4\pi)^2}$ is omitted): 

\quad- Contributions of individual volume terms in $a_4$:
\begin{equation}
\setlength{\arraycolsep}{6pt}
\begin{array}{lcl}
60RE &:&120T^2R-15R^2\nonumber\\
180E^2 &:& 720T^4 +\tfrac{45}4 R^2 -180 T^2R +180 (\nabla T)^2\nonumber\\
30\Omega^2 &:& -60T_{\mu\nu}T^{\mu\nu}-60(\nabla T)^2-120 (\nabla_\mu T_\nu)(\nabla_\nu T_\mu) 
\nonumber\\
& & \quad -\tfrac {15}4 R_{\mu\nu\rho\sigma}R^{\mu\nu\rho\sigma}
+60T^2R-120 T^{\mu}T^{\nu}R_{\mu\nu} -720 T^4 \nonumber\\
R^2\mbox{-terms} &:& 5R^2-2R_{\mu\nu}R^{\mu\nu} +2R_{\mu\nu\rho\sigma}R^{\mu\nu\rho\sigma}
\nonumber
\end{array}
\end{equation}

\quad- Contributions of individual surface terms in $a_4$:
\begin{equation}
\setlength{\arraycolsep}{6pt}
\begin{array}{lcl}
180\chi E_{;n} &: &  0\nonumber\\
30\chi R_{;n} &:&  0\nonumber\\
24L_{aa:bb} &:&  0\nonumber\\
120 EL_{aa} &:&  (240 T^2 -30R)L\nonumber\\
RL &:&  20RL+4R_{anan}L-12R_{anbn}L_{ab}
+4R_{abcb}L_{ac}\nonumber\\
L^3\mbox{-terms} &:&  \tfrac 1{21}(160L^3 -48 L_{ab}L_{ab}L+272 L_{ab}L_{bc}L_{ac})\nonumber\\
720 SE &:&  -360 LT^2 +45RL\nonumber\\
120SR &:&  -30RL\nonumber\\
144SL_{aa}L_{bb} &:&  -36L^3\nonumber\\
48SL_{ab}L_{ab} &:&  -12L_{ab}L_{ab}L\nonumber\\
480S^2L_{aa} &:&  60L^3\nonumber\\
480S^3 &:&  -30L^3\nonumber\\
120S_{:aa} &:&  0 \nonumber\\
60\chi\chi_{:a}\Omega_{an} &:&  -240T_cT_{c;n} +30L_{ab}R_{anbn}
+120LT_cT_c -120L_{ab}T_aT_b\nonumber\\
-12L\chi_{:a}\chi_{:a} &:&  -12 L_{ac}L_{ac}L -96 LT_cT_c\nonumber\\
-24L_{ab}\chi_{:a}\chi_{:b} &:&  -24L_{ab}L_{bc}L_{ac} -96LT_cT_c +96L_{ab}T_aT_b\nonumber\\
-120S\chi_{:a}\chi_{:a} &:&  30L_{ac}L_{ac}L+240LT_cT_c\nonumber
\end{array}
\end{equation}
Remark that 
\begin{align}
& \int_M 60 \,\text{tr}(E_{;\mu\mu})+\int_{\partial M} \,\text{tr} \big((240 \Pi_+-120\Pi_-)E_{;n}\big)=0, 
\label{cancelationE}    \\
& \int_M 12 \, \text{tr}(R_{;\mu\mu})+ \int_{\partial M} \text{tr} \big( (42 \Pi_+-18\Pi_-)R_{;n} \big)=0. 
\label{cancelationR}    
\end{align}
The following identity will be useful: 
\begin{eqnarray}
\int_M (\nabla T)^2=\int_M \nabla_\mu T_\nu \cdot \nabla_\nu T_\mu +R_{\mu\nu}T^\mu T^\nu
 +\int_{\partial M}  -2T_nT_{a:a} +T_n^2L +L_{ab}T_a T_b \,.
\nonumber
\end{eqnarray}
By adding up all contributions, one obtains
\begin{align}
& a_4^{(T)}=a_4^{(0)} +   \tfrac{4}{360(4\pi)^2} \biggl( \int_M   -60T_{\mu\nu} T_{\mu\nu}  \nonumber \\
&\hspace{4.3 cm} + \int_{\partial M} 48 LT_aT_a+96L_{ab}T_aT_b-240 T_a T_{a;n} 
-240 T_nT_{a:a}\,\biggr)  , \label{a4}   \\
 & a_4^{(0)}:= \tfrac{4}{360(4\pi)^2} \biggl( \int_M \tfrac{5}{4}R^2-2R_{\mu\nu}R_{\mu\nu}
-\tfrac{7}{4} R_{\mu \nu \sigma \rho}  R_{\mu \nu \sigma \rho}   \nonumber  \\
 &\hspace{0.5cm} + \int_{\partial M}  5LR +4R_{anan}L+4R_{abcb}L_{ac}+18 L_{ab}R_{anbn}
+\tfrac{34}{21} L^3+\tfrac{26}{7} LL_{ab}L_{ab}-\tfrac{232}{21} L_{ab} L_{bc} L_{ac}\, \biggr).
 \nonumber
\end{align} 
The spectral action, up to the order $\Lambda^0$, is obtained by the substitution of \eqref{a2}, \eqref{a3} and 
(\ref{a4}) in (\ref{hke}).

\subsection{Volume part of the spectral action}\label{sec-vol}

Let us discuss the volume part of the spectral action, which reads
\begin{equation}
\mathcal{S}(\DD^{(T)},\Phi)_{\rm vol}=\mathcal{S}(\DD^{(0)},\Phi)_{\rm vol}
+\int_M 8\Lambda^2 \Phi_2 T^2 -\tfrac 23 \Phi(0) T_{\mu\nu}T_{\mu\nu} 
+ \mathcal{O}(\Lambda^{-1})\, .
\label{volS}
\end{equation}
This (amazingly simple) result is fully consistent with the previous computations of $a_2, a_4$ without boundary 
on Riemann--Cartan manifolds \cite{Obukhov,Grensing,Goldthorpe,Cognola:1987wd}. 
The paper \cite{HPS} gives a much longer expression for $a_4$ containing 6 torsion-dependent terms instead of 
one in our case. Let us show that the simplification appearing in our calculations is not just a lucky 
coincidence, but rather reflects an important symmetry which has the spectral action at this order. 

Consider the identity
\begin{align}
\DD (\varphi):= e^{\varphi(x)\gamma_5}\DD^{(T)} e^{\varphi(x)\gamma_5}=
i\gamma^\mu \big(\nabla_\mu - \gamma_5 (T_\mu-\partial_\mu\varphi)\big)  \label{chiDD}
\end{align}
for an arbitrary scalar function $\varphi(x)$, which tells us that the gradient transformation of $T_\mu$ is a chiral
transformation of the Dirac operator. The corresponding infinitesimal variation of the heat trace can be computed 
by repeating the same steps as were used to derive conformal variations, see \cite{Gilkey,Gilkey2}:
\begin{align}
\frac{{\mathrm{d}}}{{\mathrm{d}}\varepsilon}\vert_{\varepsilon=0} \Tr \biggl( e^{-t\big(\DD(\varepsilon\varphi)\big)^2} 
\biggr) =-\Tr \biggl( 4t\varphi\gamma_5 \big(\DD(\varphi)\big)^2  e^{-t \big(\DD(\varphi)\big)^2}
\biggr)= 4t \frac{{\mathrm{d}}}{{\mathrm{d}}t} \Tr \biggl( \varphi\gamma_5 e^{-t \big(\DD(\varphi)\big)^2} \biggr) .
\label{dde}
\end{align}
Next, we introduce a generalization of the heat trace (\ref{hke}) "smeared" with an arbitrary endomorphism $F$ of 
the spin bundle, which also has an asymptotic expansion 
\begin{align*}
\Tr \big(Fe^{-t (\DD_{\B_\chi}(\varphi))^2}\big)  \underset{t \,\downarrow \,0}{\sim}  \, \sum_{k \in \N} 
a_{k}(F,\DD(\varphi),\chi)\, t^{(k-4)/2}\,.
\end{align*}
By expanding (\ref{dde}) in power series of $t$, we obtain the equation
\begin{align}
\frac{{\mathrm{d}}}{{\mathrm{d}}\epsilon}\vert_{\varepsilon=0}  \,(a_k)_{\rm vol}(1,\DD(\epsilon\varphi),\chi)
= 2(k-4)\,(a_k)_{\rm vol}(\varphi\gamma_5,\DD(0),\chi) 
\end{align}
yielding that the volume part of $a_4$ is invariant under the transformations (\ref{chiDD}). Consequently, only 
the terms which are invariant under $T_\mu\to T_\mu -\partial_\mu \varphi$ may appear in the volume part of 
$a_4$. The only allowed torsion dependent term is, therefore, $T_{\mu\nu}T^{\mu\nu}$. All couplings of the torsion 
to the Riemann curvature as well as the $T^4$ and $(\nabla T)^2$ terms are excluded.

These arguments cannot be used to control the boundary terms since the boundary conditions are not invariant 
under chiral transformations. The chiral invariance can be restored if one uses the chiral bag boundary 
condition, but the parameter $\theta$ should be made coordinate-dependent, see \cite{MarV} for a discussion.

When the manifold is non-compact, similar computations of coefficients $a_n^{(T)}$ can be done via a 
smearing function $f$ as in \cite{Gilkey2,V}. However, the cancellations \eqref{cancelationE}, \eqref{cancelationR} 
are not satisfied so formulae are more lengthy and, much more important, Theorem \ref{triple} has to be proved 
in this context where the spectrum of $\DD^{(T)}_\mathcal{B}$ is no more discrete.

Moreover, if $A$ is a selfadjoint one-form, i.e. $A=\sum_k a_k[\DD^{(T)}_\mathcal{B},b_k]$ with 
$a_k,b_k \in \A_{\DD^{(T)}_\B}$, then $A=i\ga_\mu A_\mu$ with $A_\mu \in C^\infty(\ol M,i\R)$ and it is possible to 
compute the fluctuations of the Dirac operator $\DD^{(T)}_\mathcal{B} +A$ applying \cite[eq. (3.26]{V}. Notice that 
a real spectral triple as no fluctuations since $\DD^{(T)}_\mathcal{B} +A+JAJ^{-1}=\DD^{(T)}_\mathcal{B}$.

\section{Stability of spectral action with respect to $\theta$}

Let us prove now that $\theta=0$ is a critical point of the spectral action which justifies the computation of the 
spectral action in section \ref{secaction}.
\begin{prop}
In any even dimension, the value $\th=0$ is a critical point for the coefficients at any order of the chiral bag spectral 
action $\SS(\DD^{(T)}_{\B_{\chi_\th}},\Phi,\Lambda)$: precisely, for any $k\in \N$,
\begin{align}
\partial_\theta |_{\theta=0} \,a_k(1,\DD^{(T)},\chi_\theta)=0.\label{da0}
\end{align}
\end{prop}

\begin{proof}
We shall need the following property
\begin{align}
\gamma_{n+1} \,\DD^{(T)} +\DD^{(T)} \, \gamma_{n+1} =0\,,\label{g5DD} 
\end{align}
which is, of course, satisfied by $\DD^{(T)}$ given in (\ref{Dirop}).

If $F$ is an auxiliary endomorphism on $\wt V$, we define the coefficients $a_k(F,\DD^{(T)},\chi_\th)$, such that the 
following complete asymptotic expansion at $t\to 0$ holds:
\begin{align*}
\Tr \big(Fe^{-t (\DD^{(T)}_{\B_{\chi_\th}})^2}\big) \underset{t \,\downarrow \,0}{\sim} \sum_{k \in \N} 
a_{k}(F,\DD^{(T)},\chi_\th)\, t^{(k-n)/2}\, .
\end{align*}
The derivative of the heat kernel coefficients with respect to $\theta$ was calculated in \cite{GK}:
\begin{align}
\partial_\theta |_{\theta=0} \,a_k(1,\DD^{(T)},\chi_\theta)=(n-k) \,a_k(\gamma_{n+1},\DD^{(T)},\chi_0) \,.
\label{datheta}
\end{align}
Let $V_1$ (resp. $V_2$) be the eigenspace of $\ga_5$ for the eigenvalue 1 (resp. $-1$) so that 
$\ol V=V_1\oplus V_2$. By \eqref{g5DD}, $\DD^{(T)}= \DD_1^{(T)} \oplus \DD_2^{(T)}$ where $\DD_1^{(T)}$ is the 
restriction of $\DD^{(T)}$ which is elliptic from $C^\infty(V_1)$ into $C^\infty(V_2)$, and $\DD_2^{(T)}$ is the 
formal adjoint of $\DD_1^{(T)}$. The mixed chiral boundary condition $\B_{\chi_0}$ can be decomposed along 
$V_1\oplus V_2$ as 
$\B_1\oplus \B_2$ so that $\DD^{(T)}_{\B_{\chi_0}}=(\DD_1^{(T)})_{\B_1}\oplus (\DD_2^{(T)})_{\B_2}$. By 
\cite[Theorem 1.9.3]{Gilkey},
\begin{align}
 \Tr \bigl(\gamma_{n+1} \, e^{-t (\DD^{(T)}_{\B_{\chi_0}})^2}\bigr)=\mbox{Index}\, \big((\DD_1^{(T)})_{\B_1} \big) 
 \quad \text{for all } t\label{Index}
\end{align}
where on the right hand side we have the index of the spin complex, i.e., the difference between the numbers of 
zero eigenmodes of $\DD^{(T)}_{\B_{\chi_0}}$ with positive and negative chirality, or more precisely 
$\dim \Ker (\DD_1^{(T)})_{\B_1} - \dim \Ker (\DD_2^{(T)})_{\B_2}$, see \cite{EGH,Gilkey}.  
By expanding both sides of this relation in asymptotic series in $t$, one obtains
\begin{align}
 &a_k(\gamma_{n+1},\DD^{(T)},\chi_0)=0 ,\quad \mbox{for} \ k\ne n.  \label{a4index}
\end{align}
Combining (\ref{a4index}) with (\ref{datheta}), one obtains the result \eqref{da0}.
\end{proof}

\begin{remark} More information can be provided about the scale invariant terms $a_n(1,\DD^{(T)},\chi_\th)$ and 
$a_n(\ga_{n+1},\DD^{(T)},\chi_0)$. First, $a_n(1,\DD^{(T)},\chi_\th)$ is actually independent of $\th$ as shown in 
\cite{GK}. Moreover, the value of $a_n(\ga_{n+1},\DD^{(T)},\chi_0)$ is given by the $t$-invariant term 
of \eqref{Index}, that is: $a_n(\ga_{n+1},\DD^{(T)},\chi_0)=\mbox{\rm Index}\,\big( (\DD_1^{(T)})_{\B_1}\big)$.
\end{remark}

\begin{remark} In dimension 4, it is actually possible to verify (\ref{a4index}) for $k\in \set{0,1,2,3}$ by using the 
formulae of \cite{EGK,MarV}. Thus, with \eqref{datheta}, we can directly get \eqref{da0} for $k\in \set{0,1,2,3,4}$. 
It is for higher coefficient orders that the index formula \eqref{Index} reveals its power.
\end{remark}

\section{Discussion}\label{sec-dis}

Let us discuss some physical implications of the spectral action we have just calculated. To achieve positive 
kinetic energies of the fields of the Standard Model, the coefficient $\Phi(0)$ has to be positive and this is the case. 
Consequently, the kinetic for $T$ seems to be negative. However, $T$ is an axial vector field rather that a vector 
field. The standard roles of the Wick rotation for such fields include an additional multiplier of $i$: an ``incorrect" 
sign of the kinetic energy in Euclidean space corresponds to a positive kinetic energy in the physical Minkowski 
case. The spectral action also predicts a mass of the torsion, which is restricted from various physical 
considerations, see \cite{Shapiro:2010zq} for a recent overview, or \cite{Shapiro:2001rz} for a more detailed 
exposition. It is considerably harder to derive physical consequences of the boundary part of the spectral action. 
The only thing we can say at the moment is that there are no linear torsion terms, so that it is less probable that 
certain boundary configurations will decay into torsion. 

Our main prediction regarding the chiral bag boundary conditions is that the point $\th=0$ is a critical point of the 
spectral action. This result is based on rather general arguments involving the Index Theorem, and therefore
is valid to all orders in $\Lambda$ and, presumably, for a larger class of spectral triples.

\section*{Acknowledgments}
We thank Thomas Sch\"ucker for useful discussions. This work was started during the visit of the third author (D. V.)
to Marseille, which was supported by the CAPES-Mathamsud program and the Universit\'e de Provence. D. V. also 
thanks CNPq and FAPESP for partial support.

\end{document}